\documentclass[
twocolumn,
superscriptaddress,
nofootinbib,
aps,
prl,
nobalancelastpage,
10pt
]{revtex4-2}
\usepackage{array}
\usepackage{amsmath}
\usepackage{amssymb}
\usepackage{bm}
\usepackage{mathtools}
\usepackage{physics}
\usepackage[dvipdfmx]{graphicx}
\usepackage{here}
\usepackage{dcolumn}
\usepackage{makecell}
\usepackage{hyperref}
\usepackage[dvipsnames]{xcolor}
\usepackage{ascmac}
\usepackage{appendix}

\hypersetup{
setpagesize=false,
bookmarksnumbered=true,
bookmarksopen=true,
colorlinks=true,
linkcolor=Blue,
citecolor=Blue,
urlcolor=magenta,
}

\begin{document}

\preprint{APS/123-QED}

\title{Thermodynamic uncertainty relation under continuous measurement and feedback with quantum-classical-transfer entropy}

\author{Kaito Tojo}
\email{
    tojo@noneq.t.u-tokyo.ac.jp
}
\affiliation{
  Department of Applied Physics, The University of Tokyo, 7-3-1 Hongo, Bunkyo-ku, Tokyo 113-8656, Japan
}

\author{Takahiro Sagawa}
\affiliation{
  Department of Applied Physics, The University of Tokyo, 7-3-1 Hongo, Bunkyo-ku, Tokyo 113-8656, Japan
}
\affiliation{
    Quantum-Phase Electronics Center (QPEC), The University of Tokyo, 7-3-1 Hongo, Bunkyo-ku, Tokyo 113-8656, Japan
}
\affiliation{
    Inamori Research Institute for Science (InaRIS), Kyoto-shi, Kyoto 600-8411, Japan
}

\author{Ken Funo}
\affiliation{
  Department of Applied Physics, The University of Tokyo, 7-3-1 Hongo, Bunkyo-ku, Tokyo 113-8656, Japan
}


\begin{abstract}
    We derive a thermodynamic uncertainty relation (TUR) under quantum continuous measurement and feedback control.
    By incorporating the quantum-classical-transfer entropy, which quantifies the information gained by continuous measurement,
    we show that the precision of currents is constrained by information-thermodynamic costs such as the entropy production and information gain.
    Our result shows that information gain has the potential to enhance the precision of currents beyond the bounds set by the conventional TUR.
    We illustrate the bound with a driven two-level system under continuous measurement and feedback, demonstrating that feedback achieves higher precision of currents while suppressing the entropy production.
\end{abstract}

\maketitle

\textit{Introduction.---}
Continuous measurement and feedback control of quantum systems are indispensable techniques for modern quantum devices~\cite{Jacobs01092006, Wiseman_Milburn_2009, ZHANG20171}.
It enables us to stabilize fragile quantum systems against dissipation or decoherence using the information obtained by measurement.
Previous studies have theoretically investigated continuous control~\cite{PhysRevLett.70.548, PhysRevA.49.2133, PhysRevA.60.2700, PhysRevLett.129.050401} and have experimentally shown that it can be utilized for state preparation~\cite{Sayrin:2011aa}, feedback cooling~\cite{Rossi:2018aa, Magrini:2021aa}, and quantum error correction~\cite{Livingston:2022aa}.

From the perspective of thermodynamics in small systems~\cite{Seifert_2012, Goold_2016,Vinjanampathy01102016,Campbell_2026_etal},
the second law of thermodynamics and the fluctuation theorem have been extended to the case with measurement and feedback~\cite{PhysRevLett.100.080403, PhysRevLett.104.090602, PhysRevE.82.061120, Toyabe:2010aa, PhysRevE.85.021104, PhysRevLett.109.180602,  PhysRevLett.111.180603, PhysRevE.88.052121, Hartich_2014, PhysRevLett.113.030601, PhysRevX.4.031015, doi:10.1073/pnas.1406966111, PhysRevA.94.012107, Ito:2016aa, PhysRevLett.117.240502, PhysRevX.7.021003, Masuyama:2018aa, PhysRevLett.122.150603, PRXQuantum.2.030353, PhysRevLett.128.170601, PhysRevLett.133.140401, prech2025quantumthermodynamicscontinuousfeedback, 5lp2-9sps} by incorporating information quantities~\cite{Parrondo:2015aa}.
As exemplified by Maxwell's demon,
measurement and feedback allows one to extract work beyond the limit set by the conventional second law.
In the quantum setting, continuous feedback enables active stabilization of the system,
opening the possibility of protocols that simultaneously achieve low dissipation and high precision of observables.
The natural question is then how these two advantages, reduced entropy production and enhanced precision, are fundamentally related.

Thermodynamic uncertainty relations, originally established in the absence of feedback control, reveal precisely such a connection in the form of a trade-off relation.
It shows that higher current precision necessarily requires larger entropy production~\cite{PhysRevLett.114.158101, PhysRevLett.116.120601, Horowitz:2020aa}.
Previous studies have extended TUR to open quantum systems~\cite{PhysRevLett.128.140602, PhysRevE.111.064107, Kwon:2025aa, PRXQuantum.6.010343} and the setting of information processing in classical~\cite{PhysRevE.100.052137, Van_Vu_2020, PhysRevLett.125.140602, PhysRevLett.125.200602, PhysRevE.101.062106, PhysRevResearch.5.043280, 4pq6-7djm} and quantum systems~\cite{f77p-kw54}.
Although some studies have derived bounds on precision of trajectory-dependent observables for quantum systems under continuous measurement and feedback using quantum dynamical activity~\cite{hasegawa2023quantumthermodynamicuncertaintyrelation, yunoki2025quantumspeedlimitquantum},
a bound based on the entropy production and information quantities for these settings remains to be elucidated.
In particular, it remains unclear how the information gained by continuous measurement can be utilized to enhance the precision of currents.

In this Letter, we derive TUR for quantum systems under continuous measurement and feedback control,
by incorporating the entropy production and the information gain.
The information gain is given by the quantum-classical (QC)- transfer entropy~\cite{PhysRevLett.128.170601}
and it quantifies the information gained by continuous measurement and is available for feedback.
The presence of these information quantities suggests that the precision of currents can be improved beyond conventional limits set by the entropy production alone for systems without measurement and feedback control.
Our TUR gives a refinement of the second law under continuous control~\cite{PhysRevLett.128.170601, 5lp2-9sps}, and clarifies the fundamental bound on the precision of currents.
We demonstrate our result by applying it to a two-level system under continuous measurement and feedback, showing that feedback successfully enhances the precision of currents.

\begin{figure}[t]
    \centering
    \includegraphics[width = 1.0\linewidth]{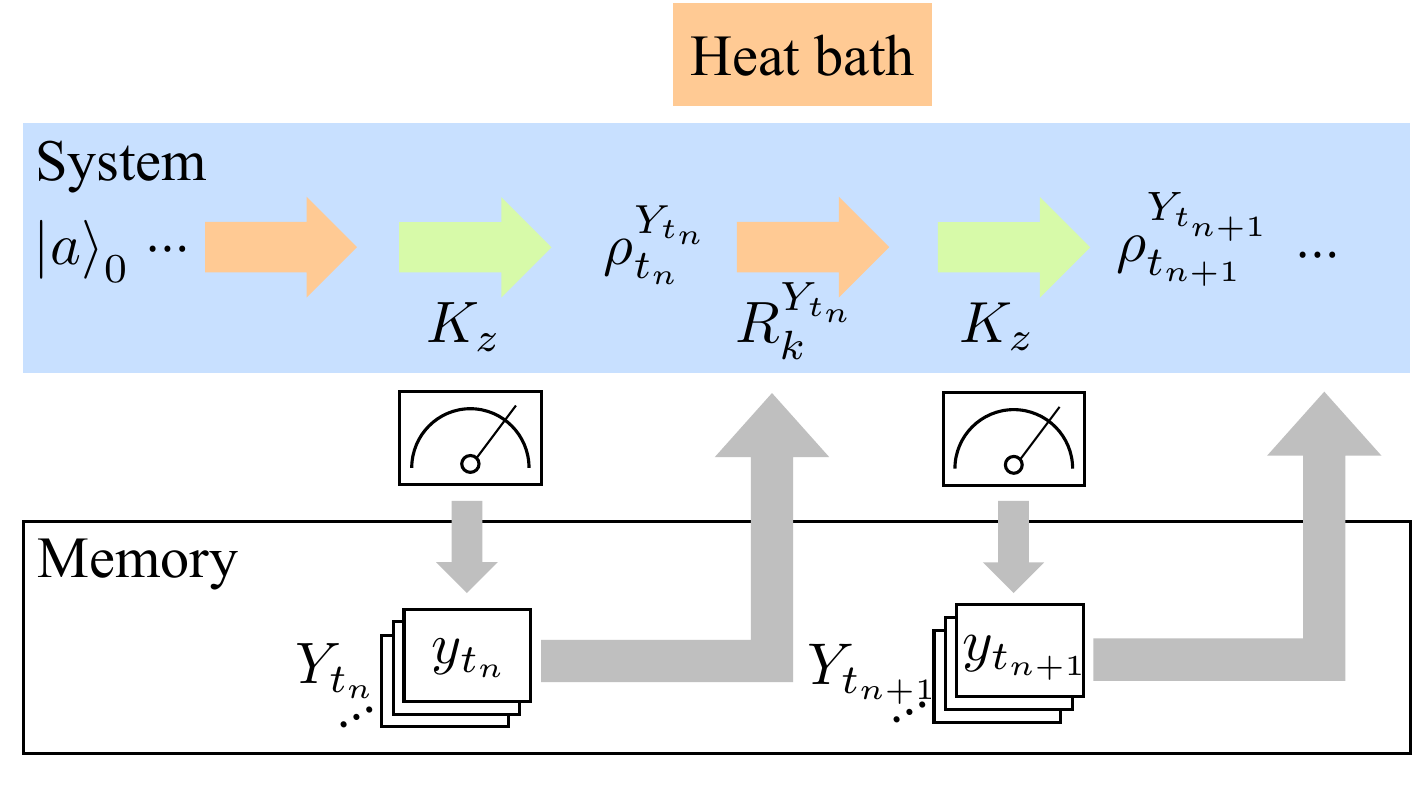}
    \caption{Schematic illustration of the stochastic trajectory of a quantum system under continuous measurement and feedback control.
    The dynamics during $[t_n, t_{n+1})$ is separated into two processes,
    interaction with the heat bath and feedback described by Kraus operators $\{R_k^{Y_{t_n}}\}_k$, and
    measurement process described by $\{K_z\}_z$.
    }
    \label{fig:schematic_dynamics}
\end{figure}

\textit{Setup.---}
\begin{figure}[b]
    \centering
    \includegraphics[width = 1.0\linewidth]{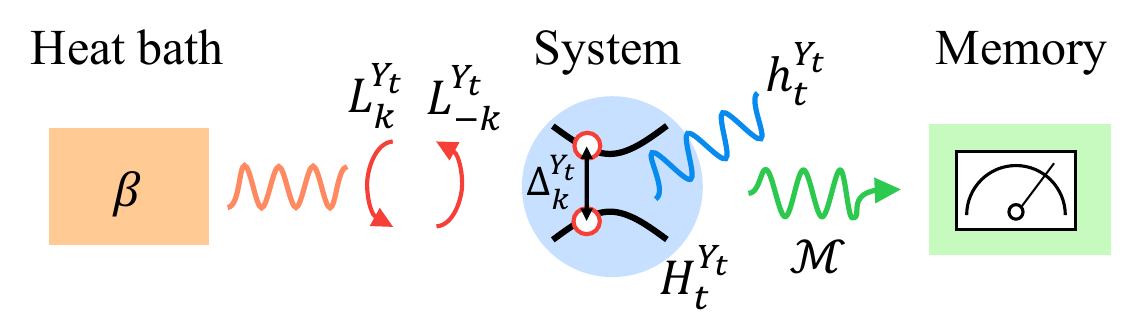}
    \caption{Schematic illustration of a quantum system attached to a heat bath under continuous measurement and feedback control.}
    \label{fig:schematic_system}
\end{figure}
We consider a quantum system attached to a heat bath at inverse temperature $\beta$ under continuous measurement and feedback
from $t = 0$ to $t = \tau$ (see Figs.~\ref{fig:schematic_system} and~\ref{fig:schematic_dynamics} for schematic illustrations).
Such situation can be described by the following stochastic master equation~\cite{10.1093/acprof:oso/9780199213900.001.0001, Albash_2012, PhysRevE.95.012136}
\begin{equation}
    \begin{split}
        &\dd \rho_{t}^{Y_{t}} = \left\{- i \left[H_{t}^{Y_{t}} + h_{t}^{Y_{t}}, \rho_{t}^{Y_{t}}\right] + \sum_{k} \mathcal{D}\left[L_{k}^{Y_{t}}\right]\rho_{t}^{Y_{t}}\right. \\
        & \left. - \sum_{z} \mathcal{H}\left[\frac{1}{2} M_{z}^\dagger M_{z}\right] \rho_{t}^{Y_{t}}\right\} \dd t + \sum_{z} \mathcal{G}\left[M_{z}\right]\rho_{t}^{Y_{t}} \cdot \dd N^{m}_{z},
    \end{split}
    \label{eq:stochastic_master_eq}
\end{equation}
where we set $\hbar = 1$.
Here, $\rho_{t}^{Y_t}$ is the quantum state of the system conditioned on measurement outcomes until time $t$, which is denoted by $Y_t$.
In Eq.~\eqref{eq:stochastic_master_eq}, we introduce the superoperators
$\mathcal{G}[M]\rho \coloneqq \frac{M \rho M^\dagger}{\Tr[M\rho M^\dagger]} - \rho$,
$\mathcal{D}[L]\rho \coloneqq L \rho L^\dagger - \frac{1}{2} \{L^\dagger L, \rho\}$,
and $\mathcal{H}[c]\rho \coloneqq c \rho + \rho c^\dagger - \Tr[c \rho + \rho c^\dagger]$.
Here, $M_z$ is the measurement operator where $z$ labels different measurement channels.
The measurement outcome during $[t, t + \dd t)$ is described by the Poisson increment $\dd N^{m}_{z}$ satisfying the property $\mathbb{E}[\dd N_{z}^{m}] = \Tr[M_z^\dagger M_z \rho_t^{Y_t}] \dd t$ and $\dd N_{z}^{m} \dd N_{z'}^{m} = \delta_{z z'} \dd N_{z}^{m}$ with $\cdot$ being It\^{o} product.
Accordingly, all measurement outcomes until time $t$ are given by $Y_t = \{\dd N^m(s)\}_{s = 0}^{t^-}$.
The system is fed back by changing the system Hamiltonian $H_t^{Y_t}$ or the driving Hamiltonian $h_{t}^{Y_t}$ based on the measurement outcomes $Y_t$ (see Fig.~\ref{fig:schematic_system}).
We assume the standard weak-coupling and Born-Markov-secular approximations for the effect of the heat bath with respect to the system Hamiltonian $H_t^{Y_t}$.
Then, the effect of the heat bath is described by the Lindblad jump operators $L_{k}^{Y_{t}}$
that dissipates heat $\Delta_{k}^{Y_{t}}$ to the heat bath and satisfies the relation $[L_k ,H_t^{Y_t}]=\Delta_k L_k^{Y_t}$.
We assume that for each $L_k^{Y_t}$, there exists its unique counterpart $L_{-k}^{Y_t}$ that describes absorption of heat $-\Delta_k^{Y_t}$ from the heat bath (see Fig.~\ref{fig:schematic_system}).
We impose the local detailed balance condition, which reads $L_{-k}^{Y_{t} \dagger} = L_{k}^{Y_{t}} e^{- \frac{\beta}{2} \Delta_{k}^{Y_{t}}}$ for all $k$.

\textit{Information thermodynamic quantities.---}
Under the setting described above, we introduce the ensemble averages of information thermodynamic quantities such as the entropy production rate $\Sigma$
and the QC-transfer entropy $I_{\mathrm{QCT}}$.

The entropy production quantifies the total entropy change of the system and the heat bath, and its rate is defined as
\begin{equation}
    \dot{\Sigma} \coloneqq \dot{S} - \beta \dot{Q},
    \label{eq:entropy_production}
\end{equation}
where $\dot{S}$ is the rate of the von Neumann entropy $S(\rho) \coloneqq - \Tr[\rho \ln \rho]$, and
$\dot{Q}$ is the rate of absorbed heat~\cite{PhysRevA.94.012107}
\begin{equation}
    \dot{Q} \coloneqq - \mathbb{E}_{Y_{t}} \left[\Tr[H_{t}^{Y_{t}} \sum_{k}\mathcal{D}[L_{k}^{Y_{t}}] \rho_{t}^{Y_{t}}]\right],
\end{equation}
where $\mathbb{E}_{Y_t}$ denotes the expectation value with respect to the measurement outcomes $Y_t$.

We now introduce the QC-transfer entropy $I_{\mathrm{QCT}}$, which quantifies the information gained by continuous measurement~\cite{PhysRevLett.128.170601}:
\begin{equation}
    \dot{I}_{\mathrm{QCT}} \coloneqq \lim_{\Delta t \to 0} \frac{1}{\Delta t} \mathbb{E}_{Y_{t_n}}\left[ I_{\mathrm{QC}}(\rho_{t}^{Y_{t}}; \dd N^m)\right].
    \label{eq:QC_transfer_entropy}
\end{equation}
Here, $I_{\mathrm{QC}}(\rho; y) \coloneqq S(\rho) - \sum_{y} p_y S\left(M_y \rho M_y^\dagger / p_y\right)$
is the QC-mutual information~\cite{Groenewold:1971aa, 10.1063/1.527179}, with $p_y \coloneqq \Tr[M_y \rho M_y^\dagger]$.
The QC-transfer entropy is the quantum extension of the transfer entropy in classical systems~\cite{PhysRevLett.85.461, PhysRevLett.111.180603}.
The generalized second law under continuous quantum measurement and feedback control is given by~\cite{PhysRevLett.128.170601}
\begin{equation}
    \Sigma + I_{\mathrm{QCT}} - \Delta \chi \geq 0,
    \label{eq:generalized_second_law}
\end{equation}
where $\Delta \chi \coloneqq \chi(\rho_{t_N}; Y_{t_N}) - \chi(\rho_{t_0}; Y_{t_0})$ denotes the change in the Holevo information between the system and measurement outcomes, where $\chi(\rho; Y) \coloneqq S(\rho) - \mathbb{E}_Y [S(\rho^Y)]$.
It is worth noting that the ordinary second law $\Sigma \geq 0$ is modified by the information quantities $I_{\mathrm{QCT}}$ and $\Delta \chi$ in the presence of continuous measurement and feedback control.
As a result, $\Sigma$ itself can be negative by acquiring information.

\textit{Measurement-feedback separation and quantum jump currents.---}
We now follow refs.~\cite{PhysRevLett.128.170601, 5cz5-n6jt} and separate the dynamics into the measurement process and the remaining part (i.e., feedback and interaction with the heat bath) to define a stochastic trajectory (see Fig.~\ref{fig:schematic_dynamics}).
To formulate this dynamics, we discretize time as $t_n \coloneqq n \Delta t$ with $n = 0, 1, \ldots, N$ and $N \Delta t = \tau$.
In this formulation, we write the measurement outcomes during $[t_n, t_{n+1})$ as $y_{t_{n+1}}$ and until $t_n$ as $Y_{t_n} \coloneqq (y_{t_1}, \ldots, y_{t_n})$ correspondingly.
The stochastic dynamics can be described by repeatedly applying the following Kraus operators in the limit of $\Delta t \to 0$ (see fig.~\ref{fig:schematic_dynamics} for schematic illustration):
\begin{equation}
    R_k^{Y_{t_n}} \coloneqq
    \begin{cases}
        L_0^{Y_{t_n}} & (k = 0)\\
        L_{k}^{Y_{t_n}} \sqrt{\Delta t}& (k \neq 0)
    \end{cases},
    \label{eq:Kraus_operators_feedback}
\end{equation}
\begin{equation}
    K_z \coloneqq
    \begin{cases}
        M_0 & (z = 0)\\
        M_z \sqrt{\Delta t} & (z \neq 0)
    \end{cases},
    \label{eq:Kraus_operators_measurement}
\end{equation}
where
$L_0^{Y_{t_n}} \coloneqq e^{- i \left(H_{t_n}^{Y_{t_n}} + h_{t_n}^{Y_{t_n}}\right)\Delta t} \left(1 - \frac{\Delta t}{2}\sum_{k} L_{k}^{Y_{t_n} \dagger} L_{k}^{Y_{t_n}}\right)$ and $M_0 \coloneqq 1 - \frac{\Delta t}{2}\sum_{z} M_z^\dagger M_z$.
That is, the dynamics during $[t_n, t_{n+1})$ is separated into two processes;
interaction with the heat bath and the time evolution generated by the Hamiltonian of the system described by $R_k^{Y_{t_n}}$,
and measurement described by $K_z$.
In this way, we can discuss the effect of measurement and feedback separately.

By monitoring all jump events during $[0, \tau]$ induced by the heat bath $K_{t_N} \coloneqq (k_{t_1}, \ldots, k_{t_N})$ and measurement $Y_{t_N}$,
and employing the two-point measurement scheme~\cite{RevModPhys.81.1665, Funo2018} using the diagonal basis of the initial state $\rho_{t_0} \coloneqq \sum_{a_0} p(a_0) \ketbra{a_0}$ and final state $\rho_\tau \coloneqq \sum_{a_\tau} p(a_\tau) \ketbra{a_\tau}$,
the stochastic trajectory is characterized by the tuple $\Gamma_\tau \coloneqq (a_0, a_\tau, K_{t_N}, Y_{t_N})$.
Accordingly, the probability distribution for observing a trajectory $\Gamma_\tau$ is given by
\begin{equation}
    p[\Gamma_\tau] = \|\bra{a_\tau}K_{y_{t_N}} R_{k_{t_N}}^{Y_{t_{N-1}}} \cdots K_{y_{t_1}} R_{k_{t_1}}^{Y_{t_0}} \ket{a_0}\|^2 p_0(a_0).
    \label{eq:probability_density_trajectory}
\end{equation}

We finally introduce quantum jump currents associated with the jump events $\{K_{t_n}\}$, which is a central quantity in this study.
Quantum jump currents are dependent on the jump events and are defined as
\begin{equation}
    \hat{J} \coloneqq \sum_{n=1}^{N} w_{k_{t_n}}^{Y_{t_n}},
\end{equation}
where $w_k^{Y_{t}}$ satisfies $w_k^{Y_{t}} = - w_{-k}^{Y_{t}}$ for all $k$ and its counterpart $-k$.
An example of such a current is the heat current, where we take $w_k^{Y_{t}} = \Delta_{k}^{Y_{t}}$.
Another example is the particle current, where we take $w_k^{Y_{t}} = \pm 1$ depending on whether the particle flows into or out of the system.
By taking the ensemble average and continuous time limit $\Delta t \to 0$, we obtain the average current
$\langle \hat{J} \rangle = \int_0^\tau \mathbb{E}_{Y_t}\left[\sum_{k} w_k^{Y_t} \Tr[L_{k}^{Y_t} \rho_{t}^{Y_t} L_{k}^{Y_t \dagger}]\right]\, \dd t$.

\textit{Main results.---}
We now present our main results, finite-time quantum TUR for general current $\hat{J}$ under continuous measurement and feedback control.
If there is no initial correlation between the system and the measurement device,
i.e., $\chi(\rho_{t_0}, Y_{t_0}) = 0$, we have
\begin{equation}
    \left(\Sigma + I_{\mathrm{QCT}} + 2 \mathcal{Q} \right)\frac{\mathrm{Var}[\hat{J}]}{\langle \hat{J}\rangle^2} \geq 2(1 + \tilde{\delta} J)^2.
    \label{eq:finite-time_TUR_no_initial_correlation}
\end{equation}
This is the simplest form of our main results
and holds for arbitrary current, initial state, and time duration of the protocol (see Appendix \hyperlink{Appendix_A}{A} for the derivation).
In Eq.~\eqref{eq:finite-time_TUR_no_initial_correlation},
we introduce the quantum correction term $Q$~\cite{PhysRevLett.128.140602} defined in Appendix \hyperlink{Appendix_B}{B},
which is shown to vanish in the classical limit or short-time limit.
We also introduce another correction term $\tilde{\delta} J$~\cite{PhysRevLett.128.140602},
which is shown to vanish in the short-time limit.
The obtained result~\eqref{eq:finite-time_TUR_no_initial_correlation} reveals the crucial role of information, along with the entropy production, in determining the precision of quantum jump currents.
In contrast to the ordinary TUR in which the entropy production $\Sigma$ alone constrain the precision of currents,
Eq.~\eqref{eq:finite-time_TUR_no_initial_correlation} shows that the QC-transfer entropy $I_{\mathrm{QCT}}$ term allows the possibility of improving the precision.
We illustrate this point by considering a two-level system under continuous measurement and feedback
and show that the precision of currents can be improved without increasing $\Sigma$ by using information.

We now make several remarks on the main result.
First, by explicitly considering initial correlation between the system and measurement device and using backward QC-transfer entropy $I_{\mathrm{BQC}}$ defined in~\cite{5lp2-9sps},
our main result~\eqref{eq:finite-time_TUR_no_initial_correlation} can be generalized to
\begin{equation}
    \begin{split}
        & \left(\Sigma + I_{\mathrm{QCT}} - \Delta \chi + 2 \mathcal{Q}\right)\frac{\mathrm{Var}[\hat{J}]}{\langle \hat{J}\rangle^2}\\
        & \geq \left(\Sigma + I_{\mathrm{QCT}} - I_{\mathrm{BQC}} + \chi(\rho_{t_0}, Y_{t_0}) + 2 \mathcal{Q} \right)\frac{\mathrm{Var}[\hat{J}]}{\langle \hat{J}\rangle^2}\\
        & \geq 2(1 + \tilde{\delta} J)^2.
    \end{split}
    \label{eq:finite-time_TUR_general}
\end{equation}
The backward QC-transfer entropy quantifies the information that is not used in the feedback process and is defined in Appendix \hyperlink{Appendix_C}{C}.
We note that when $\chi(\rho_{t_0}, Y_{t_0}) = 0$, the first inequality gives a tighter bound than Eq.~\eqref{eq:finite-time_TUR_no_initial_correlation} since $\chi(\rho_\tau; Y_\tau) \geq 0$.

Second, by taking short-time limit $\tau \to 0$, the first inequality in Eq.~\eqref{eq:finite-time_TUR_general} reduces to
\begin{equation}
    \frac{\left(\sigma + i_{\mathrm{QCT}} - x\right) \left(\tau \mathrm{Var}\left[\hat{j}\right]\right)}{\langle \hat{j} \rangle^2} \geq 2,
    \label{eq:short_time_TUR}
\end{equation}
where
$\sigma \coloneqq \Sigma/\tau,\, i_{\mathrm{QCT}} \coloneqq I_{\mathrm{QCT}}/\tau,\, x \coloneqq \Delta \chi/\tau$,\, and $\hat{j} \coloneqq \hat{J}/\tau$.
We note again that not only $\tilde{\delta} J$ but also $\mathcal{Q}$ vanishes in the short-time limit.

\begin{figure*}[ht]
    \centering
    \includegraphics[width = \linewidth]{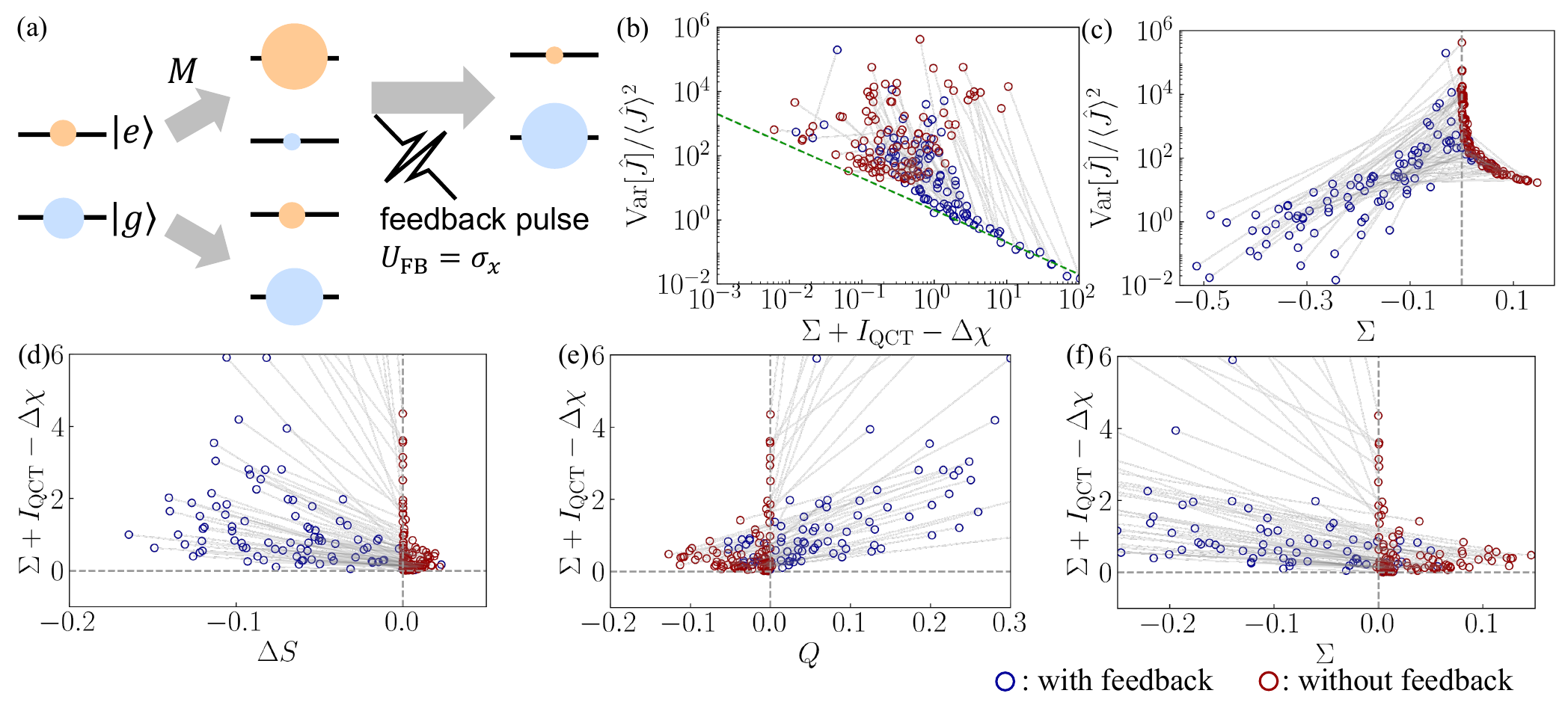}
    \caption{Numerical example of driven two-level system under continuous measurement and feedback control.
    (a) Schematic illustration of the feedback protocol.
    By applying the feedback $\pi$-pulse, we reduce the population of the excited state.
    (b)-(f) We plot the numerical results with feedback (blue circles) and without feedback (red circles) varying parameters randomly.
    Gray lines connect the points with and without feedback for the same parameters.
    (b) Scatter plot of the pairs of $\mathrm{Var}[\hat{J}]/\langle \hat{J} \rangle^2$ and $\Sigma + I_{\mathrm{QCT}} - \Delta \chi$.
    We employ the current $\hat{J}$ as the heat current from the heat bath.
    The green dotted line represents $\mathrm{Var}[\hat{J}]/\langle \hat{J} \rangle^2 = 2/(\Sigma + I_{\mathrm{QCT}} - \Delta \chi)$.
    Some points with feedback (blue circles) are below the green dotted line,
    which demonstrates that the correction terms $\mathcal{Q}, \tilde{\delta} J$ are necessary for the TUR~\eqref{eq:finite-time_TUR_no_initial_correlation} to hold.
    (c) Scatter plot of the pairs of $\mathrm{Var}[\hat{J}]/\langle \hat{J} \rangle^2$ and the entropy production $\Sigma$.
    Most points with feedback control have larger precision of the current and smaller entropy production, compared with those without feedback.
    (d) (e) (f) The scatter plots of the pairs of $\Sigma + I_{\mathrm{QCT}} - \Delta \chi$ and
    $\Delta S, Q, \Sigma$.
    The plot (d) shows that feedback successfully decreases the entropy and reduces the population of the excited state.
    The plot (f) shows that most points with feedback indicate $\Sigma < 0$, realizing Maxwell's demon.
    The plots (d), (e), (f) show that $\Sigma + I_{\mathrm{QCT}} - \Delta \chi$ is non-negative, consistent with Eq.~\eqref{eq:generalized_second_law}.
    We choose $\beta = 1, \tau = 10, \gamma_m = \gamma(2 \bar{N} + 1)$.
    We randomly sample other parameters from uniform distributions as $\omega \in [0.001, 0.5], \omega_d \in [0, 0.5], \Omega \in [0.0, 0.5], \gamma \in [0.0, 0.1], \eta \in [0.5, 1], \delta \in [0.0, 0.3]$.
    }
    \label{fig:example_numerics}
\end{figure*}

Third, we consider the classical limit such that the dynamics is described by the classical Markov jump process.
Speciffically, we consider the following conditions under which the dynamics of the system does not generate coherence between energy eigenstates:
$h_t^{Y_t} = 0$,
$H_t^{Y_t} = \sum_{i} \epsilon_t^{Y_t} \ketbra{\epsilon_i}$,
$M_{z} = \sqrt{\gamma_{m, z}} \ketbra{\epsilon_z}$, and
$\rho(0) = \sum_{i} p_0(i) \ketbra{\epsilon_i}$.
Under these conditions, the jump operators $\{L_k^{Y_t}\}$ take the form of $L_k^{Y_t} = \sqrt{\gamma_{k}^{Y_t}} \ketbra{\epsilon_{k_{\mathrm{out}}}}{\epsilon_{k_{\mathrm{in}}}}$.
Accordingly, the quantum correction term $\mathcal{Q}$~\eqref{eq:correction_term_Q} is shown to vanish as discussed in Appendix \hyperlink{Appendix_B}{B}.
Moreover, the QC-transfer entropy $I_{\mathrm{QCT}}$ reduces to the transfer entropy in classical systems~\cite{PhysRevLett.85.461, PhysRevLett.111.180603}.
The corresponding TUR of Eq.~\eqref{eq:finite-time_TUR_general} is given by
\begin{equation}
    \left(\Sigma + I_{\mathrm{TE}} - \Delta I\right)\frac{\mathrm{Var}[\hat{J}]}{\langle \hat{J}\rangle^2} \geq 2(1 + \tilde{\delta} J)^2.
    \label{eq:classical_TUR_TE}
\end{equation}
We note that by using the backward transfer entropy~\cite{Ito:2016aa},
we can also derive a tighter bound in classical limit.
While previous studies incorporate information flow and derive TUR for subsystems~\cite{PhysRevE.101.062106, PhysRevResearch.5.043280}, our result~\eqref{eq:classical_TUR_TE} incorporates the transfer entropy.

\textit{Example.---}
We numerically investigate the main result of this letter.
We employ a driven two-level system attached to a heat bath under continuous measurement and feedback control.
The system Hamiltonian and the drive Hamiltonian are given by
$H = \frac{\omega}{2} \sigma_z$ and
$h_t = \Omega (e^{- i \omega_d t} \sigma_+ + e^{i \omega_d t} \sigma_-)$,
where $\omega$ is the frequency of two-level system, and $\omega_d$ is the laser driving frequency,
and $\Omega$ is the Rabi frequency determined by the strength of the drive.
The Lindblad jump operators associated with the heat bath are given by $L_- = \sqrt{\gamma (\bar{N} + 1)} \sigma_-$ and $L_+ = \sqrt{\gamma \bar{N}} \sigma_+$,
where $\gamma$ is the dissipation rate, $\bar{N} = 1/(e^{\beta \omega} - 1)$ is the Bose-Einstein distribution,
and $\sigma_{\pm} = (\sigma_x \pm i \sigma_y)/2$.
We continuously measure the system with measurement operator
$M \coloneqq \sqrt{\gamma_m} (\ketbra{e} + \delta(\sigma_x + \ketbra{g}))$,
where $\gamma_m$ is the measurement rate and $\delta$ represents the measurement error.
When the measurement outcome indicates that the system is in the excited state,
we apply a feedback $\pi$-pulse represented by the unitary operator $U_{\mathrm{FB}} = e^{- i h' \Delta t} = \sigma_x$ to reduce the population of the excited state and cool the system (see Fig.~\ref{fig:example_numerics} (a)).
We note that this feedback protocol does not use the past measurement outcomes and is thus Markovian.
We also introduce the imperfection of the detection with efficiency $0 < \eta \leq 1$.
In other words, we introduce two types of jump operators $\sqrt{\eta} M$ and $\sqrt{1 - \eta} M$ corresponding to detected and undetected measurement outcomes, respectively.
We take the initial state of the system as the Gibbs state with respect to the system Hamiltonian.
We numerically calculate the change of von Neumann entropy $\Delta S$,
the ensemble average of the heat current absorbed from the heat bath $Q$,
the entropy production $\Sigma = \Delta S - \beta Q$,
$\Sigma + I_{\mathrm{QCT}} - \Delta \chi$,
and the variance of the heat current $\mathrm{Var}[\hat{J}]$.
Figure~\ref{fig:example_numerics} shows that feedback successfully decreases the entropy production of the system $\Sigma$, thereby realizing Maxwell's demon.
It also shows that feedback control typically achieves higher precision of the current without increasing the entropy production $\Sigma$.

\textit{Conclusion.---}
We established a quantum thermodynamic uncertainty relation in the setting of continuous measurement and feedback control~\eqref{eq:finite-time_TUR_no_initial_correlation}.
Our main results show that the precision of currents is constrained not only by the entropy production $\Sigma$ but also by the QC-transfer entropy $I_{\mathrm{QCT}}$, which quantifies the information obtained by continuous measurement.
While the TURs without measurement and feedback implies that the entropy production alone limits the precision of currents, our results show that we can further improve its precision by utilizing the obtained information via feedback control.
Our numerical example of two-level systems shows that feedback control typically achieves both higher precision of currents and lower entropy production.

We briefly discuss some future directions.
Beyond the Markovian feedback protocol analyzed in our example, our framework applies to more general feedback strategies, including non-Markovian schemes in which the control depends on the full measurement history.
Investigating non-Markovian features of the feedback and its implication on the TUR is a future issue.
Another future direction would be to derive other types of bounds on precision of trajectory observables, such as the kinetic uncertainty relation or first-passage time~\cite{PhysRevE.95.032134, Di_Terlizzi_2019, PhysRevE.105.044127}.

\textit{Note added.---} Independent of our work, the authors of Ref.~\cite{honma2026thermodynamicuncertaintyrelationquantum} obtained related but distinct results for quantum TUR under feedback control by incorporating the quantum mutual information.

\textit{Acknowledgments.---}
We thank Isaac Layton for fruitful discussions.
We also thank Ryotaro Honma and Tan Van Vu for sharing their preliminary manuscripts.
The numerical calculations were done by using the QuTiP library~\cite{lambert2025qutip5quantumtoolbox}.
This work is supported by JST ERATO Grant No. JPMJER2302, Japan.
K.T. is supported by World-leading Innovative Graduate Study Program for Materials Research, Information, and Technology (MERIT-WINGS) of the University of Tokyo.
T.S. is supported by JST CREST Grant No. JPMJCR20C1 and by Institute of AI and Beyond of the University of Tokyo.
K.F. is supported by JSPS KAKENHI Grant Nos. JP23K13036 and JP24H00831.

\textit{Data availability.---}
The data are not publicly available upon publication. The data are available from the au-
thors upon reasonable request.
\bibliography{ref}

\section*{End matter}

\hypertarget{Appendix_A}{\textit{Appendix A: Derivation of the main result using QC-transfer entropy.---}}
We derive quantum TUR for general current $\hat{J}$ under continuous measurement and feedback control by incorporating QC-transfer entropy.
In analogy with the previous study~\cite{PhysRevLett.128.140602},
we consider the auxiliary perturbation of the Hamiltonian and jump operators $\{L_k^{Y_t}\}$ using parameter $\theta$ as
\begin{equation}
    H_{t, \theta}^{Y_t} \coloneqq (1 + \theta) H_{t}^{Y_t}, \quad h_{t, \theta}^{Y_t} \coloneqq (1 + \theta) h_{t}^{Y_t},
    \label{eq:perturbation_Hamiltonian}
\end{equation}
\begin{equation}
    L_{k, \theta}^{Y_t} = \sqrt{1 + \ell_{k}^{Y_t} \theta} L_{k}^{Y_{t}},
    \label{eq:perturbation_lindblad}
\end{equation}
where
\begin{equation}
    \ell_{k}^{Y_t} \coloneqq \frac{\Tr[L_{k}^{Y_{t}} \rho_{t}^{Y_{t}} L_{k}^{Y_{t} \dagger}] - \Tr[L_{-k}^{Y_{t}} \rho_{t}^{Y_{t}} L_{-k}^{Y_{t} \dagger}]}{\Tr[L_{k}^{Y_{t}} \rho_{t}^{Y_{t}} L_{k}^{Y_{t} \dagger}] + \Tr[L_{-k}^{Y_{t}} \rho_{t}^{Y_{t}} L_{-k}^{Y_{t} \dagger}]}.
\end{equation}
Note that $\ell_k^{Y_t} = - \ell_{-k}^{Y_t}$ and $\ell_0^{Y_t} = 0$.
We also note that in contrast to previous studies~\cite{PhysRevLett.128.140602}, only the jump operators related to the heat bath are perturbed as shown in Eq.~\eqref{eq:perturbation_lindblad}.
These perturbations~\eqref{eq:perturbation_Hamiltonian} and~\eqref{eq:perturbation_lindblad} are analogous to the derivation of TUR for subsystems~\cite{PhysRevResearch.5.043280, f77p-kw54}.
Accordingly, Kraus operators are perturbed as
\begin{equation}
    R_{k, \theta}^{Y_{t}} \coloneqq
    \begin{cases}
        \tilde{L}_{0, \theta}^{Y_{t}} & (k = 0)\\
        L_{k, \theta}^{Y_{t}} \sqrt{\Delta t}& (k \neq 0)
    \end{cases},
\end{equation}
where
\begin{equation}
    \begin{split}
        \tilde{L}_{0, \theta}^{Y_{t}} &\coloneqq \exp\left[- i \left((1+\theta)(H_{t}^{Y_t} + h_{t}^{Y_t})\right)\Delta t\right]\\
                                  & \cdot \left(1 - \frac{\Delta t}{2}\sum_{k} L_{k, \theta}^{Y_{t} \dagger} L_{k, \theta}^{Y_{t}}\right).
    \end{split}
\end{equation}
The probability density of realizing a trajectory $\Gamma_\tau = (a_0, a_\tau, K_{t_N}, Y_{t_N})$ under the perturbed dynamics is given by

\begin{equation}
    \begin{split}
        p_\theta[\Gamma_\tau] &= \|\bra{a_\tau}K_{y_{t_N}} R_{k_{t_N}, \theta}^{Y_{t_{N-1}}} \cdots K_{y_{t_1}} R_{k_{t_1}, \theta}^{Y_{t_0}}\ket{a_0}\|^2 p_0(a_0)\\
                              &= \Pi_{n=1}^{N} (1 + \ell_{k_{t_n}}^{Y_{t_{n-1}}} \theta)\\
                              &\phantom{=} \cdot \|\bra{a_\tau}K_{y_{t_N}} R_{k_{t_N}, \theta}^{' Y_{t_{N-1}}} \cdots K_{y_{t_1}} R_{k_{t_1}, \theta}^{' Y_{t_0}}\ket{a_0}\|^2 p_0(a_0),
    \end{split}
    \label{eq:probability_density_parameterized}
\end{equation}
where
\begin{equation}
    R_{k, \theta}^{' Y_{t}} \coloneqq
    \begin{cases}
        \tilde{L}_{0, \theta}^{Y_{t}} & (k = 0)\\
        L_{k}^{Y_{t}} \sqrt{\Delta t}& (k \neq 0)
    \end{cases}.
\end{equation}

To derive main results, we consider estimating parameter $\theta$ by measuring the quantum jump current $\hat{J}$~\cite{PhysRevLett.128.140602}.
Using Cram\'er-Rao inequality, the precision of the current $\hat{J}$ at $\theta = 0$ is bounded as
\begin{equation}
    \frac{\mathrm{Var}[\hat{J}]}{\left.\left(\partial_\theta \langle \hat{J}\rangle_{\theta} \right|_{\theta = 0}\right)^2} \geq \frac{1}{\mathcal{I}(0)},
    \label{eq:cramer_rao}
\end{equation}
where $\langle \cdot \rangle_\theta$ denotes expectation value with respect to  $p_\theta[\Gamma_\tau]$ and $\mathcal{I}(\theta)$ is the Fisher information defined as
\begin{equation}
    \mathcal{I}(\theta) \coloneqq - \expval{\pdv[2]{\theta} \ln p_\theta[\Gamma_\tau]}_\theta.
\end{equation}
The partial derivative of the average current at $\theta = 0$ appearing in Eq.~\eqref{eq:cramer_rao} is calculated by expanding the conditional state $\rho_{t, \theta}^{Y_t}$ as $\rho_{t, \theta}^{Y_t} = \rho_t^{Y_t} + \theta \phi_t^{Y_t} + o(\theta)$ and is given by
\begin{equation}
    \begin{split}
        \partial_\theta\left.\langle \hat{J}\rangle_{\theta} \right|_{\theta = 0} &= \pdv{\theta} \int_0^\tau \mathbb{E}_{Y_t}\left[\sum_{k} \Tr[L_{k}^{Y_{t}} (\rho_t^{Y_t} + \theta \phi_t^{Y_t}) L_{k}^{Y_{t} \dagger}]\right.\\
    &\phantom{=} \left. \left. \cdot w_k^{Y_{t}} (1 + \ell_{k}^{Y_{t}} \theta) + o(\theta)\right] \dd t \right|_{\theta = 0}\\
    &= \langle \hat{J} \rangle + \langle \hat{J}\rangle_*\\
    &= (1 + \tilde{\delta} J) \langle \hat{J}\rangle,
    \end{split}
    \label{eq:current_derivative_theta}
\end{equation}
where $\langle \hat{J} \rangle_* \coloneqq \int_0^\tau \mathbb{E}_{Y_t}\left[\sum_{k} w_k \left(\Tr[L_{k}^{Y_{t}} \phi_t^{Y_t} L_{k}^{Y_{t} \dagger}]\right) \right]\dd t$
, and $\tilde{\delta} J \coloneqq \frac{\langle \hat{J}\rangle_*}{\langle \hat{J}\rangle}$ denotes the correction term.

For parameterized probability of the trajectory $\Gamma_\tau$~\eqref{eq:probability_density_parameterized},
the Fisher information is bounded from above as
\begin{equation}
    \mathcal{I}(0) \leq \frac{1}{2} \left(\Sigma + I_{\mathrm{QCT}} - \Delta \chi\right) + \mathcal{Q},
    \label{eq:bound_fisher_information}
\end{equation}
where $\mathcal{Q}$ is given by
\begin{align}
    \begin{split}
        &\mathcal{Q} =\\
        &- \left.\expval{\pdv[2]{\theta} \ln \|\bra{a_\tau}K_{y_{t_N}} R_{k_{t_N}, \theta}^{' Y_{t_{N-1}}} \cdots K_{y_{t_1}} R_{k_{t_1}, \theta}^{' Y_{t_0}}\ket{a_0}\|^2}_{\theta}\right|_{\theta = 0}.
    \end{split}
    \label{eq:correction_term_Q}
\end{align}
The derivation of Eq.~\eqref{eq:bound_fisher_information} is provided in supplementary material~\cite{supplementary}.

We are now in a position to derive the quantum TUR under continuous measurement and feedback control.
Substituting Eqs.~\eqref{eq:current_derivative_theta} and~\eqref{eq:bound_fisher_information} to the Cram\'{e}r-Rao inequality~\eqref{eq:cramer_rao}, we have
\begin{equation}
    \left(\Sigma + I_{\mathrm{QCT}} - \Delta \chi + 2\mathcal{Q} \right)\frac{\mathrm{Var}[\hat{J}]}{\langle \hat{J}\rangle^2} \geq 2(1 + \tilde{\delta} J)^2.
    \label{eq:finite-time_TUR}
\end{equation}
If there is no initial correlation between the system and measurement device, and using the non-negativity of Holevo information $\chi(\rho_\tau; Y_\tau) \geq 0$,
we have
\begin{equation}
    \left(\Sigma + I_{\mathrm{QCT}} + 2 \mathcal{Q} \right)\frac{\mathrm{Var}[\hat{J}]}{\langle \hat{J}\rangle^2} \geq 2(1 + \tilde{\delta} J)^2,
\end{equation}
which is also shown in the main text as Eq.~\eqref{eq:finite-time_TUR_no_initial_correlation}.
We note that while our derivation is based on Cram\'er-Rao inequality for classical Fisher information,
recent studies have derived quantum TURs using quantum Fisher information~\cite{PRXQuantum.6.010343, f77p-kw54}.
In these studies, the correction term $\tilde{\delta} J$ appears, although its form is different.
Also, the correction term $\mathcal{Q}$ does not appear.
It would be an interesting future work to investigate the quantum TUR under continuous measurement and feedback control using quantum Fisher information.

\hypertarget{Appendix_B}{\textit{Appendix B: On the correction terms.---}}
We discuss the correction terms $\mathcal{Q}$ and $\tilde{\delta} J$ in our finite-time quantum TUR~\eqref{eq:finite-time_TUR_no_initial_correlation}.
We first provide further details of the classical limit discussed in the main text.
That is, the quantum correction term $\mathcal{Q}$~\eqref{eq:correction_term_Q} is shown to be reduced to the following form:
\begin{equation}
    \mathcal{Q} = - 2 \left. \left\langle \pdv[2]{\theta} \sum_{n = 1}^N \delta_{k_{t_n}, 0} \ln \abs{1 + \theta \Delta t g_n} \right\rangle \right|_{\theta = 0} = \mathcal{O}(\Delta t),
    \label{eq:quantum_correction_classical}
\end{equation}
where
\begin{equation}
    g_n = - \frac{\ell_{n_{\mathrm{in}}}^{Y_{t_{n-1}}} \gamma_{n_{\mathrm{in}}}^{Y_{t_{n-1}}}}{2\left(1 - \frac{\Delta t \gamma_{n_{\mathrm{in}}}^{Y_{t_{n-1}}}}{2}\right)}- \frac{\ell_{n_{\mathrm{in}}}^{Y_{t_{n-1}}} \gamma_{n_{\mathrm{in}}}^{Y_{t_{n-1}}}}{2} + o(\Delta t).
\end{equation}
$g_n$ is a trajectory-dependent quantity and $\mathcal{O}(1)$ in $\Delta t$.
Its details are provided in the supplementary material~\cite{supplementary}.

We next consider the short-time limit of the finite-time TUR~\eqref{eq:finite-time_TUR} for $[t, t+\Delta t]$.
We can show that the $\mathcal{Q}$ term is $\mathcal{O}(\Delta t^2)$, which vanishes in the short-time limit (see supplementary material~\cite{supplementary} for details).
Also, due to the initial condition $\phi_t^{Y_t} = O$, we have $\tilde{\delta} J =\mathbb{E}_{Y_t}[\sum_k w_k^{Y_t} L_k^{Y_t} \phi_{t}^{Y_{t}} L_{k}^{Y_t \dagger}] \Delta t = 0$.
Therefore, we have the short-time TUR given in Eq.~\eqref{eq:short_time_TUR}.
We note that it is possible to derive the short-time TUR directly by applying the Cauchy-Schwarz inequality in analogy with classical systems~\cite{PhysRevE.101.062106}.

\hypertarget{Appendix_C}{\textit{Appendix C: Backward QC-transfer entropy and extension of the main result.---}}
Here, we introduce the backward QC-transfer entropy $I_{\mathrm{BQC}}$~\cite{5lp2-9sps}, which quantifies the information that is not used in the feedback process,
to provide a tighter bound than Eq.~\eqref{eq:finite-time_TUR}.
To explicitly discuss the information that is not utilized for feedback, we consider for now the situation where the feedback on time $t_n$ is based on the partial measurement outcomes
from $t_n - t_m (n \geq m)$ to $t_n$, i.e.,
$Y_{t_n}^{t_m} \coloneqq \{y_{t_{n-m+1}}, \ldots, y_{t_n}\}$.
We take the limit $\Delta t \to 0$ keeping $t_m = m \Delta t \eqcolon \tau_{\mathrm{fb}}$ finite.
In this case, backward QC-transfer entropy is given by~\cite{5lp2-9sps}
\begin{equation}
    \begin{split}
        I_{\mathrm{BQC}} &\coloneqq \sum_{n = m+1}^{N-1} \left(\chi(\rho_{t_n}; Y_{t_{n-m}} | Y_{t_n}^{t_m}) - \chi(\sigma_{t_n}; Y_{t_{n-m}} | Y_{t_n}^{t_m})\right)\\
                     &\phantom{\coloneqq} + \chi(\rho_{t_N}; Y_{t_N})\\
                     &\geq 0,
    \end{split}
    \label{eq:backward_QC_transfer_entropy}
\end{equation}
where conditional Holevo information is defined as $\chi(\rho; Y|Z) \coloneqq \mathbb{E}_{Z}[S(\rho^Z)] - \mathbb{E}_{Y, Z}[S(\rho^{Y, Z})]$.
Since not all measurement outcomes are used in the feedback process at time $t_n$,
backward QC-transfer entropy quantifies the information that is not utilized in the feedback process.
The corresponding generalized second law using backward QC-transfer entropy is given by~\cite{5lp2-9sps}
\begin{equation}
    \Sigma + I_{\mathrm{QCT}} - I_{\mathrm{BQC}} + \chi(\rho_{t_0}; Y_{t_0}) \geq 0,
    \label{eq:generalized_second_law_backward_QC_transfer_entropy}
\end{equation}
which is tighter than Eq.~\eqref{eq:generalized_second_law} since $I_{\mathrm{BQC}} \geq \chi(\rho_{t_N}; Y_{t_N})$
(see supplemental material~\cite{supplementary} for further details.).

By explicitly considering the structure of feedback control, we extend the finite-time TUR~\eqref{eq:finite-time_TUR}.
If the feedback at time $t_n$ is based on the partial measurement outcomes $Y_{t_n}^{\tau_{\mathrm{fb}}}$,
the time evolution of the conditional state $\rho_t^{Y_t^{\tau_{\mathrm{fb}}}}$ is given by replacing $Y_t$ in Eq.~\eqref{eq:stochastic_master_eq} with $Y_t^{\tau_{\mathrm{fb}}}$.
Accordingly, by replacing $Y_t$ with $Y_t^{\tau_{\mathrm{fb}}}$ in the above derivation of quantum TUR using QC-transfer entropy,
we obtain the quantum TUR using backward QC-transfer entropy as
\begin{equation}
    \begin{split}
        &\left(\Sigma + I_{\mathrm{QCT}} - I_{\mathrm{BQC}} + \chi(\rho_{t_0}; Y_{t_0}) + 2\mathcal{Q} \right)\frac{\mathrm{Var}[\hat{J}]}{\langle \hat{J}\rangle^2}\\
        &\geq 2(1 + \tilde{\delta} J)^2.
    \end{split}
    \label{eq:finite_time_TUR_backward_QC_transfer_entropy}
\end{equation}
We note that Eq.~\eqref{eq:finite_time_TUR_backward_QC_transfer_entropy} is tighter than Eq.~\eqref{eq:finite-time_TUR} since $I_{\mathrm{BQC}} \geq \Delta \chi$.


\end{document}


\preprint{APS/123-QED}

\title{Supplementary Material for ``Thermodynamic uncertainty relation under continuous measurement and feedback with quantum-classical-transfer entropy''}

\author{Kaito Tojo}
\email{
    tojo@noneq.t.u-tokyo.ac.jp
}
\affiliation{
  Department of Applied Physics, The University of Tokyo, 7-3-1 Hongo, Bunkyo-ku, Tokyo 113-8656, Japan
}

\author{Takahiro Sagawa}
\affiliation{
  Department of Applied Physics, The University of Tokyo, 7-3-1 Hongo, Bunkyo-ku, Tokyo 113-8656, Japan
}
\affiliation{
    Quantum-Phase Electronics Center (QPEC), The University of Tokyo, 7-3-1 Hongo, Bunkyo-ku, Tokyo 113-8656, Japan
}
\affiliation{
    Inamori Research Institute for Science (InaRIS), Kyoto-shi, Kyoto 600-8411, Japan
}

\author{Ken Funo}
\affiliation{
  Department of Applied Physics, The University of Tokyo, 7-3-1 Hongo, Bunkyo-ku, Tokyo 113-8656, Japan
}


\maketitle

The supplementary material is organized as follows.
In Sec.~\ref{sec:quantum_information_flow_and_generalized_second_law},
we provide further details of information-thermodynamic quantities such as the QC-transfer entropy and backward QC-transfer entropy,
and generalized second law presented in the main text.
In Sec.~\ref{sec:derivation_main_results}, we provide further details of the derivation of main results presented in the End Matter.
In Sec.~\ref{sec:correction_term}, we provide further details of the correction term $\mathcal{Q}$~(25) in the short-time limit and the classical limit.
In Sec.~\ref{sec:details_numerical_example}, we provide details of the numerical example.

\section{Quantum Information Flow and Generalized Second Law}\label{sec:quantum_information_flow_and_generalized_second_law}
\begin{figure}[b]
    \centering
    \includegraphics[width = 0.6\linewidth]{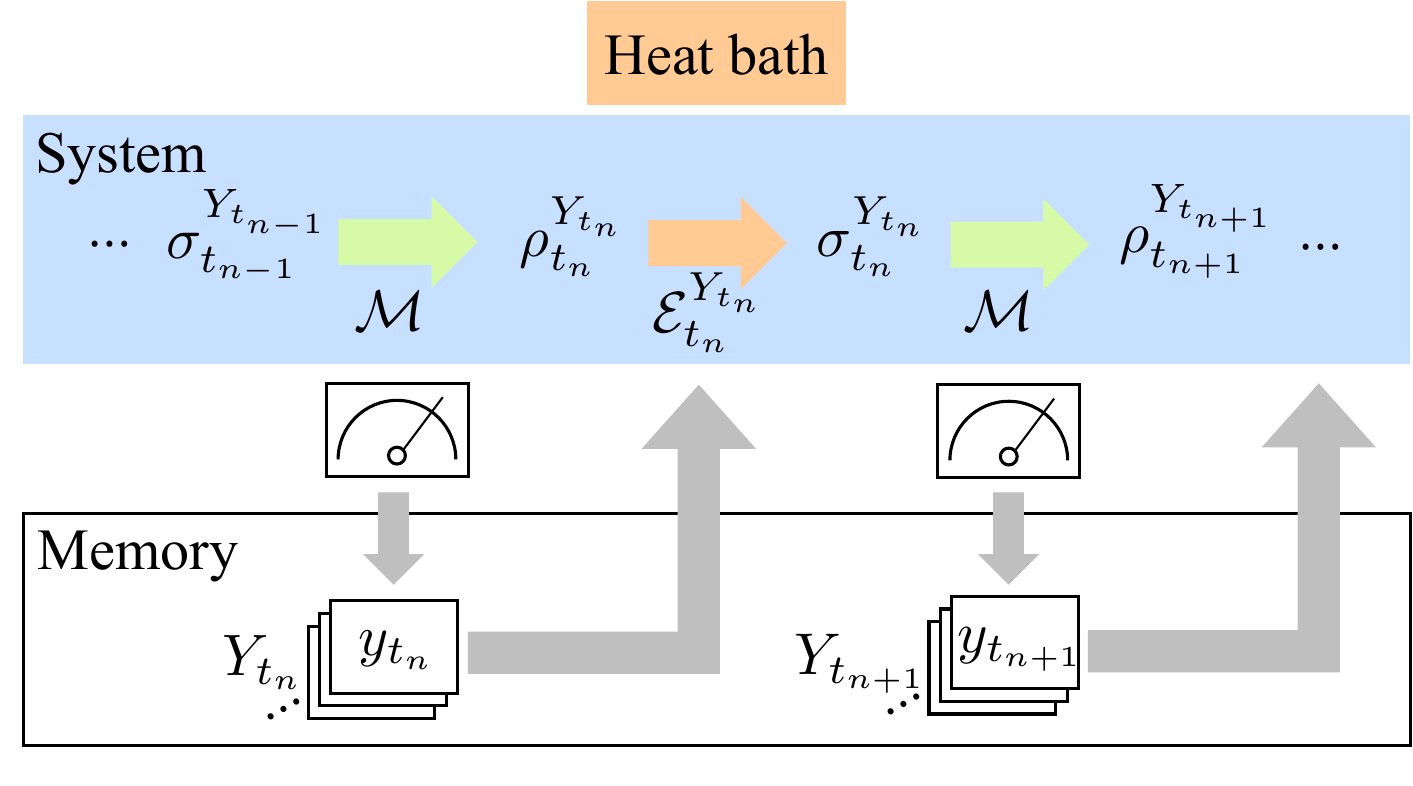}
    \caption{Schematic illustration of the dynamics of a quantum system under continuous measurement and feedback control.
    $Y_{t_n} \coloneqq (y_{t_1}, y_{t_2}, \ldots,  y_{t_n})$ denotes accumulated measurement outcomes up to $t_n$.
    The dynamics during $[t_n, t_{n+1})$ is separated into two processes,
    interaction with the heat bath and feedback denoted by $\mathcal{E}_{t_{n}}^{Y_{n}}$, and
    measurement process denoted by $\mathcal{M}$.
    The states right before and after the process $\mathcal{E}_{t_{n}}^{Y_{n}}$ are denoted by $\rho_{t_{n}}^{Y_{t_n}}$ and $\sigma_{t_n}^{Y_{t_n}}$.}
    \label{fig:schematic_dynamics}
\end{figure}
In this section, we provide further details of information-thermodynamic quantities and generalized second law presented in the main text.
In particular, we use the expression Eq.~\eqref{eq:generalized_second_law_QC_transfer_entropy_continuous_limit} of the generalized second law when deriving the main result, as presented in Appendix 2.

We consider the setup of a quantum system under continuous measurement and feedback control, where the dynamics is separated into two processes during $[t_n, t_{n+1})$ as shown in Fig.~\ref{fig:schematic_dynamics}.
We introduce CPTP maps representing these two processes as $\mathcal{E}_{t_n}^{Y_{t_n}}(\rho) \coloneqq \sum_{k} R_k^{Y_{t_n}} \rho R_k^{Y_{t_n} \dagger}$ and $\mathcal{M}(\sigma) \coloneqq \sum_{z} K_z \sigma K_z^\dagger$.
We also introduce the states right before and after the feedback process during $[t_n, t_{n+1})$ conditioned on $Y_{t_n}$ as $\rho_{t_n}^{Y_n}$ and $\sigma_{t_n}^{Y_{t_n}}$.

\subsection{The generalized second law using QC-transfer entropy}
Here, we review the derivation of the generalized second law using QC-transfer entropy~\cite{PhysRevLett.128.170601, 5lp2-9sps, 5cz5-n6jt}, which is formulated as $\Sigma + I_{\mathrm{QCT}} - \Delta \chi \geq 0$.
The QC-transfer entropy is defined as
\begin{equation}
    I_{\mathrm{QCT}} \coloneqq \sum_{n=0}^{N-1} \mathbb{E}_{Y_{t_n}}\left[ I_{\mathrm{QC}}(\sigma_{t_n}^{Y_{t_n}}; y_{t_{n+1}})\right].
    \label{eq:QC_transfer_entropy}
\end{equation}
We note that in the limit of $\Delta t \to 0$, the QC-transfer entropy rate is reduced to Eq.~(4) in the main text since $\sigma_{t_n}^{Y_{t_n}} = \rho_{t_n}^{Y_{t_n}} + \mathcal{O}(\Delta t)$.
By denoting the Gibbs state with respect to the inverse temperature $\beta$ and the system Hamiltonian $H$ as $\rho_\beta(H) \coloneqq e^{-\beta H}/\Tr[e^{-\beta H}]$,
$\Sigma + I_{\mathrm{QCT}} - \Delta \chi$ is rewritten as
\begin{equation}
    \begin{split}
        \Sigma + I_{\mathrm{QCT}} - \Delta \chi &= S(\rho_\tau) - S(\rho_{t_0}) + \beta \sum_{n = 0}^{N-1} \mathbb{E}_{Y_{t_n}}\left[\Tr\left[H_{t_n}^{Y_{t_n}}\sum_{k} \mathcal{D}[L_k^{Y_{t_n}}]\rho_{t_{n}}^{Y_{t_{n}}}\right]\right] \Delta t \\
                                            &\phantom{=} + \sum_{n = 0}^{N-1} \mathbb{E}_{Y_{t_{n}}}\left[S(\sigma_{t_n}^{Y_{t_n}}) - \mathbb{E}_{y_{t_{n+1}}}[\rho_{t_{n+1}}^{Y_{t_{n+1}}}]\right] - \left(S(\rho_{t_N}) - \mathbb{E}_{Y_{t_N}}[S(\rho_{t_N}^{Y_{t_N}})]\right) + \left(S(\rho_{t_0}) - \mathbb{E}_{Y_{t_0}}[S(\rho_{t_0}^{Y_{t_0}})]\right)\\
                                            &= \sum_{n=0}^{N-1} \mathbb{E}_{Y_{t_n}}\left[S(\sigma_{t_n}^{Y_{t_n}}) - S(\rho_{t_n}^{Y_{t_n}}) - \Tr[\ln \rho_{\beta}(H_{t_n}^{Y_{t_n}}) \sum_k \mathcal{D}[L_k^{Y_{t_n}}] \rho_{t_n}^{Y_{t_n}}]\right].
    \end{split}
    \label{eq:generalized_second_law_QC_transfer_entropy_proof_first}
\end{equation}
By defining the CPTP map
\begin{equation}
    \tilde{\mathcal{E}}_{t_n}^{Y_{t_n}}(\rho) \coloneqq \left(1 - \frac{\Delta t}{2} \sum_k L_k^{Y_{t_n} \dagger} L_k^{Y_{t_n}}\right) \rho \left(1 - \frac{\Delta t}{2} \sum_k L_k^{Y_{t_n} \dagger} L_k^{Y_{t_n}}\right) + \Delta t \sum_k L_k^{Y_{t_n}} \rho L_k^{Y_{t_n} \dagger},
\end{equation}
we can show the following relations hold up to the first order of $\Delta t$:
\begin{equation}
    S(\sigma_{t_n}^{Y_{t_n}}) = S(\mathcal{E}_{t_n}^{Y_{t_n}}(\rho_{t_n}^{Y_{t_n}})) = S(\tilde{\mathcal{E}}_{t_n}^{Y_{t_n}}(\rho_{t_n}^{Y_{t_n}})) + o(\Delta t)
\end{equation}
and
\begin{equation}
    \sum_k \mathcal{D}[L_k^{Y_{t_n}}]\rho_{t_n}^{Y_{t_n}} = \tilde{\mathcal{E}}_{t_n}^{Y_{t_n}}(\rho_{t_n}^{Y_{t_n}}) - \rho_{t_n}^{Y_{t_n}} + o(\Delta t).
\end{equation}
Thus, we have
\begin{equation}
    \begin{split}
        \Sigma + I_{\mathrm{QCT}} - \Delta \chi &= \sum_{n=0}^{N-1} \mathbb{E}_{Y_{t_n}}\left[- \Tr[\tilde{\mathcal{E}}_{t_n}^{Y_{t_n}}(\rho_{t_n}^{Y_{t_n}}) \ln \tilde{\mathcal{E}}_{t_n}^{Y_{t_n}}(\rho_{t_n}^{Y_{t_n}})] - S(\rho_{t_n}^{Y_{t_n}}) + \Tr[\left(\tilde{\mathcal{E}}_{t_n}^{Y_{t_n}}(\rho_{t_n}^{Y_{t_n}}) - \rho_{t_n}^{Y_{t_n}}\right) \ln \rho_\beta(H_{t_n}^{Y_{t_n}})]\right]\\
                                                &= \sum_{n=0}^{N-1} \mathbb{E}_{Y_{t_n}}\left[S(\rho_{t_n}^{Y_{t_n}} \| \rho_\beta(H_{t_n}^{Y_{t_n}})) - S\left(\tilde{\mathcal{E}}_{t_n}^{Y_{t_n}}(\rho_{t_n}^{Y_{t_n}}) || \tilde{\mathcal{E}}_{t_n}^{Y_{t_n}}(\rho_\beta(H_{t_n}^{Y_{t_n}}))\right)\right] \geq 0,
    \end{split}
    \label{eq:generalized_second_law_QC_transfer_entropy_proof_second}
\end{equation}
where we used $\mathcal{\tilde{E}}_{t_n}^{Y_{t_n}}(\rho_\beta(H_{t_n}^{Y_{t_n}})) = \rho_\beta(H_{t_n}^{Y_{t_n}})$ and the monotonicity of relative entropy under CPTP maps.

Furthermore, by denoting the spectral decomposition of the state $\rho_{t_n}^{Y_{t_n}}$ as
\begin{equation}
    \rho_{t_n}^{Y_{t_n}} = \sum_{c_n} p_{t_n}^{Y_{t_n}}(c_n) \ketbra{c_n},
\end{equation}
and defining the transition rate of the eigenstates of $\rho_{t_n}^{Y_{t_n}}$ as
\begin{equation}
    W_{k, c_n' c_n}^{Y_{t_n}} \coloneqq \abs{\bra{c_n'} L_{k}^{Y_{t_n}} \ket{c_n}}^2,
\end{equation}
the generalized second law~\eqref{eq:generalized_second_law_QC_transfer_entropy_proof_second} is expressed as
\begin{equation}
    \begin{split}
        \Sigma + I_{\mathrm{QCT}} - \Delta \chi &= \frac{1}{2} \sum_{n=0}^{N-1} \Delta t \mathbb{E}_{Y_{t_n}}\left[\sum_{k \neq 0, c_n, c_n'} \left(W_{c_n', c_n}^{k, Y_{t_n}} p_{t_n}^{Y_{t_n}}(c_n) - W_{c_n, c_n'}^{-k, Y_{t_n}} p_{t_n}^{Y_{t_n}}(c_n')\right) \cdot \ln \frac{W_{c_n', c_n}^{k, Y_{t_n}} p_{t_n}^{Y_{t_n}}(c_n)}{W_{c_n, c_n'}^{-k, Y_{t_n}} p_{t_n}^{Y_{t_n}} (c_n')}\right]\\
                                                &\xrightarrow{\substack{\Delta t \to 0,\\ N \Delta t = \tau}} \frac{1}{2} \int_0^\tau \mathbb{E}_{Y_t}\left[\sum_{k \neq 0, c_t, c_t'} \left(W_{c_t', c_t}^{k, Y_t} p_{t}^{Y_t}(c_t) - W_{c_t, c_t'}^{-k, Y_t} p_{t}^{Y_t}(c_t')\right) \cdot \ln \frac{W_{c_t', c_t}^{k, Y_t} p_{t}^{Y_t}(c_t)}{W_{c_t, c_t'}^{-k, Y_t} p_{t}^{Y_t} (c_t')}\right] \dd t \geq 0.
    \end{split}
    \label{eq:generalized_second_law_QC_transfer_entropy_continuous_limit}
\end{equation}

\subsection{The generalized second law using the backward QC-transfer entropy and its properties}
In this section, we review some properties of the backward QC-transfer entropy and the derivation of the generalized second law using the backward QC-transfer entropy~\cite{5lp2-9sps}.
We first show the backward QC-transfer entropy is bounded below by Holevo information $\chi(\rho_\tau; Y_\tau)$, which is nonnegative.
The backward QC-transfer entropy~(30) is rewritten as
\begin{equation}
    \begin{split}
        I_{\mathrm{BQC}} &= \sum_{n = m+1}^{N-1} \left(\chi(\rho_{t_n}; Y_{t_{n-m}} | Y_{t_n}^{t_m}) - \chi(\sigma_{t_n}; Y_{t_{n-m}} | Y_{t_n}^{t_m})\right) + \chi(\rho_{t_N}; Y_{t_N})\\
                         &= \sum_{n = m+1}^{N-1} \mathbb{E}_{Y_{t_n}} \left[\left(S(\rho_{t_n}^{Y_{t_n}^{t_m}}) - S(\rho_{t_n}^{Y_{t_n}})\right) - \left(S(\sigma_{t_n}^{Y_{t_n}^{t_m}}) - S(\sigma_{t_n}^{Y_{t_n}})\right)\right] + \chi(\rho_{t_N}; Y_{t_N})\\
                         &= \sum_{n = m+1}^{N-1} \mathbb{E}_{Y_{t_n}} \left[\tr[- \rho_{t_n}^{Y_{t_n}} \left(\ln \rho_{t_n}^{Y_{t_n}^{t_m}} - \ln \rho_{t_n}^{Y_{t_n}}\right)] - \tr[- \sigma_{t_n}^{Y_{t_n}} \left(\ln \sigma_{t_n}^{Y_{t_n}^{t_m}} - \ln \sigma_{t_n}^{Y_{t_n}}\right)]\right] + \chi(\rho_{t_N}; Y_{t_N})\\
                         &= \sum_{n = m+1}^{N-1} \mathbb{E}_{Y_{t_n}} \left[S(\rho_{t_n}^{Y_{t_n}} || \rho_{t_n}^{Y_{t_n}^{t_m}}) - S(\mathcal{E}_{t_n}^{Y_{t_n}^{t_m}} (\rho_{t_n}^{Y_{t_n}}) || \mathcal{E}_{t_n}^{Y_{t_n}^{t_m}} (\rho_{t_n}^{Y_{t_n}^{t_m}}))\right] + \chi(\rho_{t_N}; Y_{t_N})\\
                         &\geq \chi(\rho_{t_N}; Y_{t_N}) \geq 0,
    \end{split}
\end{equation}
where we used the monotonicity of relative entropy under CPTP maps and the nonnegativity of the Holevo information.

Next, we show the generalized second law using the backward QC-transfer entropy (31).
In a similar manner to the derivation of Eqs.~\eqref{eq:generalized_second_law_QC_transfer_entropy_proof_first} and~\eqref{eq:generalized_second_law_QC_transfer_entropy_proof_second}, the generalized second law using backward QC-transfer entropy is shown as
\begin{equation}
    \begin{split}
        \Sigma + I_{\mathrm{QCT}} - I_{\mathrm{BQC}} + \chi(\rho_{t_0}; Y_{t_0}) &= \sum_{n=0}^{N-1} \mathbb{E}_{Y_{t_n}}\left[S(\sigma_{t_n}^{Y_{t_n}}) - S(\rho_{t_n}^{Y_{t_n}}) - \Tr[\ln \rho_{\beta}(H_{t_n}^{Y_{t_n}^{t_m}}) \sum_k \mathcal{D}[L_k^{Y_{t_n}^{t_m}}] \rho_{t_n}^{Y_{t_n}^{t_m}}]\right] \\
                                                                                 &\phantom{=} - \sum_{n = m+1}^{N-1} \mathbb{E}_{Y_{t_n}} \left[\left(S(\rho_{t_n}^{Y_{t_n}^{t_m}}) - S(\rho_{t_n}^{Y_{t_n}})\right) - \left(S(\sigma_{t_n}^{Y_{t_n}^{t_m}}) - S(\sigma_{t_n}^{Y_{t_n}})\right)\right] \\
                                                                                 &= \sum_{n=0}^{N-1} \mathbb{E}_{Y_{t_n}^{t_m}}\left[S(\sigma_{t_n}^{Y_{t_n}^{t_m}}) - S(\rho_{t_n}^{Y_{t_n}^{t_m}}) - \Tr[\ln \rho_{\beta}(H_{t_n}^{Y_{t_n}^{t_m}}) \sum_k \mathcal{D}[L_k^{Y_{t_n}^{t_m}}] \rho_{t_n}^{Y_{t_n}^{t_m}}]\right]\\
                                                                                 &= \sum_{n=0}^{N-1} \mathbb{E}_{Y_{t_n}^{t_m}}\left[S(\rho_{t_n}^{Y_{t_n}^{t_m}} \| \rho_\beta(H_{t_n}^{Y_{t_n}^{t_m}})) - S\left(\tilde{\mathcal{E}}_{t_n}^{Y_{t_n}^{t_m}}(\rho_{t_n}^{Y_{t_n}^{t_m}}) || \tilde{\mathcal{E}}_{t_n}^{Y_{t_n}^{t_m}}(\rho_\beta(H_{t_n}^{Y_{t_n}^{t_m}}))\right)\right]\\
                                                                                 &\geq 0,
    \end{split}
\end{equation}
where the CPTP map $\tilde{\mathcal{E}}_{t_n}^{Y_{t_n}^{t_m}}$ is defined replacing $L_k^{Y_{t_n}}$ with $L_k^{Y_{t_n}^{t_m}}$ in the definition of $\tilde{\mathcal{E}}_{t_n}^{Y_{t_n}}$ and the inequality is due to the monotonicity of relative entropy under CPTP maps.

In the same way as Eq.~\eqref{eq:generalized_second_law_QC_transfer_entropy_continuous_limit}, the generalized second law using the backward QC-transfer entropy is expressed in the continuous-time limit as
\begin{align}
    \begin{split}
        & \Sigma + I_{\mathrm{QCT}} - I_{\mathrm{BQC}} + \chi(\rho_{t_0}; Y_{t_0}) \\
        &\xrightarrow{\substack{\Delta t \to 0,\\ N \Delta t = \tau, m \Delta t = \tau_{\mathrm{fb}}}} \frac{1}{2} \int_0^\tau \mathbb{E}_{Y_t^{\tau_{\mathrm{fb}}}}\left[\sum_{k \neq 0, c_t, c_t'} \left(W_{c_t', c_t}^{k, Y_t^{\tau_{\mathrm{fb}}}} p_{t}^{Y_t^{\tau_{\mathrm{fb}}}}(c_t) - W_{c_t, c_t'}^{-k, Y_t^{\tau_{\mathrm{fb}}}} p_{t}^{Y_t^{\tau_{\mathrm{fb}}}}(c_t')\right) \cdot \ln \frac{W_{c_t', c_t}^{k, Y_t^{\tau_{\mathrm{fb}}}} p_{t}^{Y_t^{\tau_{\mathrm{fb}}}}(c_t)}{W_{c_t, c_t'}^{-k, Y_t^{\tau_{\mathrm{fb}}}} p_{t}^{Y_t^{\tau_{\mathrm{fb}}}} (c_t')}\right] \dd t \geq 0,
    \end{split}
    \label{eq:generalized_second_law_backward_QC_transfer_entropy_continuous_limit}
\end{align}
We note that Eq.~\eqref{eq:generalized_second_law_backward_QC_transfer_entropy_continuous_limit} is reduced to  Eq.~\eqref{eq:generalized_second_law_QC_transfer_entropy_continuous_limit} when $\tau_{\mathrm{fb}} = 0$.

\section{Details of the derivation of main results}\label{sec:derivation_main_results}
In this section, we provide further details of the derivation of main results presented in the End Matter.
Specifically, we show that the Fisher information for the parameterized probability density of the trajectory $\Gamma_\tau$~(19) at $\theta = 0$
is bounded from above by the entropy production, the QC-transfer entropy, the difference in Holevo information and the correction term~(25).
Here, we provide its derivation.

The Fisher information at $\theta = 0$ is given by
\begin{equation}
    \begin{split}
        \mathcal{I}(0) &= - \left. \expval{\pdv[2]{\theta} \ln p_\theta[\Gamma_\tau]}_\theta\right|_{\theta = 0} \\
                       &= - \left. \expval{\pdv[2]{\theta} \ln \prod_{n = 1}^{N} (1 + \ell_{k_{t_n}}^{Y_{t_{n-1}}} \theta)}_\theta\right|_{\theta = 0} - \left. \expval{\pdv[2]{\theta} \ln \|\bra{a_\tau}K_{y_{t_N}} R_{k_{t_N}, \theta}^{' Y_{t_{N-1}}} \cdots K_{y_{t_1}} R_{k_{t_1}, \theta}^{' Y_{t_0}}\ket{a_0}\|^2}_\theta\right|_{\theta = 0}\\
                       &= \left. \expval{\sum_{n = 1}^{N} \left(\ell_{k_{t_n}}^{Y_{t_{n-1}}}\right)^2}_\theta\right|_{\theta = 0} + \mathcal{Q}.
    \end{split}
    \label{eq:Fisher_information_expression}
\end{equation}
Taking the continuous-time limit, the first term is calculated as
\begin{equation}
    \begin{split}
        \lim_{\substack{\Delta t \to 0, \\ N \Delta t = \tau}} \left. \expval{\sum_{n = 1}^{N} \left(\ell_{k_{t_n}}^{Y_{t_{n-1}}}\right)^2}_{\theta} \right|_{\theta = 0} &= \int_0^\tau \mathbb{E}_{Y_t} \left[\sum_{k} \Tr[L_{k}^{Y_{t}} \rho_t^{Y_t} L_{k}^{Y_{t} \dagger}] \left(\ell_{k}^{Y_t}\right)^2\right] \dd t\\
                                                                                                                                                                          &= \frac{1}{2} \int_0^\tau \mathbb{E}_{Y_t} \left[\sum_{k} \left(\Tr[L_{k}^{Y_{t}} \rho_t^{Y_t} L_{k}^{Y_{t} \dagger}] + \Tr[L_{-k}^{Y_{t}} \rho_t^{Y_t} L_{-k}^{Y_{t} \dagger}]\right) \left(\ell_{k}^{Y_t}\right)^2\right] \dd t\\
                                                                                                                                                                          &= \frac{1}{2}\int_0^\tau \mathbb{E}_{Y_t} \left[\sum_{k \neq 0} \frac{\left(\Tr[L_{k}^{Y_{t}} \rho_t^{Y_t} L_{k}^{Y_{t} \dagger}] - \Tr[L_{-k}^{Y_{t}} \rho_t^{Y_t} L_{-k}^{Y_{t} \dagger}]\right)^2}{\Tr[L_{k}^{Y_{t}} \rho_t^{Y_t} L_{k}^{Y_{t} \dagger}] + \Tr[L_{-k}^{Y_{t}} \rho_t^{Y_t} L_{-k}^{Y_{t} \dagger}]}\right] \dd t\\
                                                                                                                                                                          &= \frac{1}{2}\int_0^\tau \mathbb{E}_{Y_t} \left[\sum_{k\neq 0} \frac{\left\{\sum_{c_t, c_t'} \left(p_t^{Y_t}(c_t) W_{k, c_t' c_t}^{Y_t} - p_t^{Y_t}(c_t') W_{k, c_t c_t'}^{Y_t}\right)^2\right\}}{\sum_{c_{t}, c_t'} p_t^{Y_t}(c_t) W_{k, c_t' c_t}^{Y_t} + p_t^{Y_t}(c_t') W_{k, c_t c_t'}^{Y_t}}\right] \dd t\\
                                                                                                                                                                          &\leq \frac{1}{2}\int_0^\tau \mathbb{E}_{Y_t} \left[\sum_{k \neq 0} \sum_{c_t, c_t'} \frac{\left(p_t^{Y_t}(c_t) W_{k, c_t' c_t}^{Y_t} - p_t^{Y_t}(c_t') W_{k, c_t c_t'}^{Y_t}\right)^2}{p_t^{Y_t}(c_t) W_{k, c_t' c_t}^{Y_t} + p_t^{Y_t}(c_t') W_{k, c_t c_t'}^{Y_t}}\right] \dd t\\
                                                                                                                                                                          &\leq \frac{1}{4}\int_0^\tau \mathbb{E}_{Y_t} \left[\sum_{k \neq 0} \sum_{c_t, c_t'} \left(p_t^{Y_t}(c_t) W_{k, c_t' c_t}^{Y_t} - p_t^{Y_t}(c_t') W_{k, c_t c_t'}^{Y_t}\right) \ln \frac{p_t^{Y_t}(c_t) W_{k, c_t' c_t}^{Y_t}}{p_t^{Y_t}(c_t') W_{k, c_t c_t'}^{Y_t}}\right] \dd t\\
                                                                                                                                                                          &= \frac{1}{2} \left(\Sigma + I_{\mathrm{QCT}} - \Delta \chi\right),
    \end{split}
    \label{eq:bound_partial_EP_QC_transfer_entropy}
\end{equation}
where we used Cauchy-Schwarz inequality $\sum_i (a_i + b_i) \sum_j \frac{(a_j - b_j)^2}{a_j + b_j} \geq \left\{\sum_i (a_i - b_i)\right\}^2$ for $a_i, b_i \geq 0$ and the inequality $(a-b) \ln \frac{a}{b} \geq 2 \frac{(a-b)^2}{a+b}$ for $a, b \geq 0$.
Using Eqs.~\eqref{eq:Fisher_information_expression} and~\eqref{eq:bound_partial_EP_QC_transfer_entropy}, we obtain the upper bound on the Fisher information in the main text.

Since Eq.~\eqref{eq:generalized_second_law_backward_QC_transfer_entropy_continuous_limit} is obtained by formally replacing $Y_t$ with $Y_t^{\tau_{\mathrm{fb}}}$ in Eq.~\eqref{eq:generalized_second_law_QC_transfer_entropy_continuous_limit},
we can also obtain another upper bound on the Fisher information for the feedback structure introduced in the End Matter as
\begin{equation}
    \mathcal{I}(0) \leq \frac{1}{2} \left(\Sigma + I_{\mathrm{QCT}} - I_{\mathrm{BQC}} + \chi(\rho_{t_0}; Y_{t_0})\right) + \mathcal{Q}.
    \label{eq:bound_fisher_information_backward_QC_transfer_entropy}
\end{equation}

\section{Details of the correction term}\label{sec:correction_term}
In this section, we provide further details of the correction term $\mathcal{Q}$~(25) appearing in the upper bound on the Fisher information~(24).

\subsection{Correction term in the short-time limit}\label{appendix_subsec:correction_term_in_short_time_limit}
In this section, we show that the correction term $\mathcal{Q}$~(25) vanishes in the short-time limit.
In the short-time limit, the $\mathcal{Q}$ term is given by
\begin{equation}
    \begin{split}
        \mathcal{Q} &= - \left.\expval{\pdv[2]{\theta} \ln \|\bra{a_{t+\Delta t}}K_{y_{t+\Delta t}} R_{k_t, \theta}^{' Y_{t}}\ket{a_t}\|^2}_{\theta}\right|_{\theta = 0}\\
                    &= - \left.\expval{\pdv[2]{\theta} \ln \|\bra{a_{t+\Delta t}}K_{y_{t+\Delta t}} \tilde{L}_{0, \theta}^{Y_t}\delta_{k_t, 0} \ket{a_t}\|^2}_{\theta}\right|_{\theta = 0}.
    \end{split}
    \label{eq:evaluation_Q_term}
\end{equation}
By expanding $\tilde{L}_{0, \theta}^{Y_t}$ up to the second order of $\theta$, we have
\begin{equation}
    \begin{split}
        \tilde{L}_{0, \theta}^{Y_t} &= \exp\left[-i (1 + \theta)\left(H_t^{Y_t} + h_t^{Y_t}\right) \Delta t\right] \left(1 - \frac{\Delta t}{2} \sum_k (1 + \theta \ell_k^{Y_t})L_k^{Y_t \dagger} L_k^{Y_t}\right) \\
                                    &= \exp\left[- i \left(H_t^{Y_t} + h_t^{Y_t}\right) \Delta t\right]\left[1 - \frac{\Delta t}{2}\sum_k L_k^{Y_t \dagger} L_k^{Y_t}\right. \\
                                    &\phantom{=} \left. + \theta \Delta t \left\{- i \left(H_t^{Y_t} + h_t^{Y_t}\right)\left(1 - \frac{\Delta t}{2} \sum_k L_k^{Y_t \dagger} L_k^{Y_t} \right) - \frac{1}{2}\sum_k \ell_k^{Y_t} L_k^{Y_t\dagger} L_k^{Y_t}\right\}\right. \\
                                    &\phantom{=} \left. + \theta^2 \Delta t^2 \left\{- \frac{1}{2} \left(H_t^{Y_t} + h_t^{Y_t}\right)^2 \left(1 - \frac{\Delta t}{2}\sum_k L_k^{Y_t\dagger} L_k^{Y_t}\right) + \frac{i}{2} \left(H_t^{Y_t} + h_t^{Y_t}\right) \sum_k \ell_k^{Y_t} L_k^{Y_t\dagger} L_k^{Y_t}\right\}\right]+ \mathcal{O}(\Delta t^3).
    \end{split}
    \label{eq:expansion_tilde_L_0}
\end{equation}
Eq.~\eqref{eq:expansion_tilde_L_0} implies that $\pdv[2]\theta \ln \|\bra{a_{t+\Delta t}} K_{y_t} \tilde{L}_{0, \theta}^{Y_t} \ket{a_t}\|^2$ appearing in Eq.~\eqref{eq:evaluation_Q_term} can be expressed as
\begin{equation}
    \left. \left\langle \pdv[2]\theta \ln \|\bra{a_{t+\Delta t}} K_{y_t} \tilde{L}_{0, \theta}^{Y_t} \ket{a_t}\|^2 \right\rangle_\theta \right|_{\theta = 0} = \left. \pdv[2]{\theta} \ln (1 + f(\theta)) \right|_{\theta = 0},
\end{equation}
where $f(\theta)$ is a function satisfying $f(0) = 0$, $f'(\theta) = \mathcal{O}(\Delta t)$, and $f''(0) = \mathcal{O}(\Delta t^2)$.
Since
\begin{equation}
    \left. \pdv[2]{\theta} \ln (1 + f(\theta))\right|_{\theta = 0} = f''(0) - (f'(0))^2 = \mathcal{O}(\Delta t^2),
\end{equation}
we conclude that the correction term $\mathcal{Q}$ vanishes in the short-time limit.

\subsection{Correction term in the classical limit}\label{subsec:correction_term_in_classical_limit}
In this section, we show that the correction term $\mathcal{Q}$ reduces to the form~(28) in the classical limit.
Under the conditions of the classical limit mentioned in the main text, the application of Kraus operators $K_{y_{t_n}} R^{' Y_{t_n}}_{k_{t_n}, \theta}$ to the energy eigenstate $\ket{\epsilon_{n_{\mathrm{in}}}}$ is proportional to another energy eigenstate as
\begin{equation}
    \begin{split}
        K_{y_{t_n}} R^{' Y_{t_{n-1}}}_{k_{t_n}, \theta} \ket{\epsilon_{n_{\mathrm{in}}}} &= \left[\delta_{y_{t_n}, 0} \left(1 - \frac{\Delta t}{2} \sum_{z} \gamma_{m, z} \ketbra{\epsilon_z}\right) + (1 - \delta_{y_{t_n}, 0}) \sqrt{\gamma_{m, y_{t_n}} \Delta t} \ketbra{\epsilon_{y_{t_n}}}\right]\\
                                                                                         &\phantom{=} \cdot \left[\delta_{k_{t_n}, 0} \exp\left\{-i(1+\theta) H_{t_{n}}^{Y_{t_{n-1}}} \Delta t\right\} \left(1 - \frac{\Delta t}{2} \sum_{k} (1 + \ell_k^{Y_{t_{n-1}}} \theta)L_k^{Y_{t_{n-1}}\dagger} L_k^{Y_{t_{n-1}}}\right)\right.\\
                                                                                         &\phantom{=} \left. + (1 - \delta_{k_{t_n}, 0}) \sqrt{\gamma_{k_{t_n}}^{Y_{t_{n-1}}} \Delta t} \ketbra{\epsilon_{k_{t_n, \mathrm{out}}}}{\epsilon_{k_{t_n, \mathrm{in}}}}\right] \ket{\epsilon_{n_{\mathrm{in}}}} \\
                                                                                         &= \delta_{k_{t_n}, 0} \left[\delta_{y_{t_n}, 0} \left(1 - \frac{\Delta t}{2} \gamma_{m, n_{\mathrm{in}}} \right) + (1 - \delta_{y_{t_n}, 0}) \delta_{y_{t_n}, n_{\mathrm{in}}} \sqrt{\gamma_{m, y_{t_n}} \Delta t}\right] \\
                                                                                         &\phantom{=} \cdot \exp\left\{-i(1+\theta) \epsilon_{n_{\mathrm{in}}} \Delta t\right\} \left(1 - \frac{\Delta t}{2} (1 + \ell_{n_{\mathrm{in}}}^{Y_{t_{n-1}}} \theta) \gamma_{n_{\mathrm{in}}}^{Y_{t_{n-1}}}\right) \ket{\epsilon_{n_{\mathrm{in}}}}\\
                                                                                         &\phantom{=} + (1 - \delta_{k_{t_n}, 0}) \delta_{k_{t_n, \mathrm{in}}, n_{\mathrm{in}}} \left[\delta_{y_{t_n}, 0} \left(1 - \frac{\Delta t}{2} \gamma_{m, t_{n, \mathrm{in}}} \right) + (1 - \delta_{y_{t_n}, 0}) \delta_{y_{t_n}, t_{n, \mathrm{in}}} \sqrt{\gamma_{m, y_{t_n}} \Delta t}\right] \\
                                                                                         &\phantom{=} \cdot \sqrt{\gamma_{k_{t_n}}^{Y_{t_{n-1}}} \Delta t} \ket{\epsilon_{k_{t_n}, \mathrm{out}}}.
    \end{split}
\end{equation}
Thus, the $\theta$-dependent terms in the correction term $\mathcal{Q}$~(25) are factorized and expressed as
\begin{equation}
    \begin{split}
        - \pdv[2]{\theta} \ln \|\bra{a_\tau}K_{y_{t_N}} R_{k_{t_N}, \theta}^{' Y_{t_{N-1}}} \cdots K_{y_{t_1}} R_{k_{t_1}, \theta}^{' Y_{t_0}}\ket{a_0}\|^2 &= - 2 \pdv[2]{\theta} \sum_{n = 1}^{N} \delta_{k_{t_n}, 0} \ln \abs{1 - \frac{\Delta t}{2} (1 + \ell_{n_{\mathrm{in}}}^{Y_{t_{n-1}}} \theta) \gamma_{n_{\mathrm{in}}}^{Y_{t_{n-1}}}}\\
                                                                                                                                                            &= - 2 \pdv[2]{\theta} \sum_{n = 1}^{N} \delta_{k_{t_n}, 0} \ln \abs{1 - \frac{\ell_{n_{\mathrm{in}}}^{Y_{t_{n-1}}} \gamma_{n_{\mathrm{in}}}^{Y_{t_{n-1}}}}{2\left(1 - \frac{\Delta t \gamma_{n_{\mathrm{in}}}^{Y_{t_{n-1}}}}{2}\right)} \theta \Delta t}.
    \end{split}
\end{equation}
We simplify the above expression by defining
\begin{equation}
    g_n \coloneqq - \frac{\ell_{n_{\mathrm{in}}}^{Y_{t_{n-1}}} \gamma_{n_{\mathrm{in}}}^{Y_{t_{n-1}}}}{2\left(1 - \frac{\Delta t \gamma_{n_{\mathrm{in}}}^{Y_{t_{n-1}}}}{2}\right)} = - \frac{\ell_{n_{\mathrm{in}}}^{Y_{t_{n-1}}} \gamma_{n_{\mathrm{in}}}^{Y_{t_{n-1}}}}{2} + o(\Delta t),
\end{equation}
and obtain the expression of the correction term $\mathcal{Q}$ in the classical limit in Eq.~(28).

\section{Details of the numerical example}\label{sec:details_numerical_example}
In the main text, we present a numerical example of two-level system under continuous measurement and feedback control.
We calculate $10^4$ trajectories of stochastic Schrödinger equation~\cite{10.1093/acprof:oso/9780199213900.001.0001, Wiseman_Milburn_2009}
\begin{equation}
    \begin{split}
        &\dd \ket*{\psi_{t}^{Y_{t}}} = \left\{- i (H_{t}^{Y_{t}} + h_{t}^{Y_{t}}) - \frac{1}{2} \sum_{k} \left(L_k^{Y_{t}\dagger} L_k^{Y_{t}} - \|L_k^{Y_{t}} \ket*{\psi_t^{Y_t}}\|^2 \right) - \frac{1}{2}\sum_{z} \left(M_z^\dagger M_z - \|M_z \ket*{\psi_t^{Y_t}}\|^2 \right) \right\} \ket*{\psi_t^{Y_t}} \dd t \\
        &+ \sum_{k} \left(\frac{L_k^{Y_t} \ket*{\psi_t^{Y_t}}}{\|L_k^{Y_t} \ket*{\psi_t^{Y_t}}\|} - \ket*{\psi_t^{Y_t}}\right) \cdot \dd N_k^b + \sum_z \left(\frac{M_z \ket*{\psi_t^{Y_t}}}{\|M_z \ket*{\psi_t^{Y_t}}\|} - \ket*{\psi_t^{Y_t}}\right) \cdot \dd N_z^m
    \end{split}
    \label{eq:stochastic_Schrodinger_eq}
\end{equation}
for each parameters to obtain the change of von Neumann entropy $\Delta S$, ensemble average and variance of the heat current $\langle \hat{J} \rangle, \mathrm{Var}[\hat{J}]$.
Here, $\{\dd N_k^b\}$ are the Poisson increments corresponding to the quantum jumps induced by the Lindblad operators $\{L_k^{Y_t}\}$.
We also calculate $10^3$ trajectories of stochastic master equation~(1) using quantum jump unraveling with the measurement operator $M$ to obtain $\rho_t^{Y_t}$, with which we calculate $\Sigma + I_{\mathrm{QCT}} - \Delta \chi$.
The numerical calculations were done by using the QuTiP library~\cite{lambert2025qutip5quantumtoolbox}.

\bibliography{ref_sm}